\documentclass[iop,apjl]{emulateapj}
\usepackage{apjfonts}

\usepackage{acronym}

\newacro{GRB}[GRB]{$\gamma$\nobreakdashes-ray burst}
\newacro{GBM}[GBM]{Gamma-ray Burst Monitor}
\newacro{IPN}[IPN]{InterPlanetary Network}
\newacro{LAT}[LAT]{Large Area Telescope}
\newacro{XRT}[XRT]{X-Ray Telescope}
\newacro{CARMA}[CARMA]{Combined Array for Research in Millimeter-wave Astronomy}
\newacro{GW}[GW]{gravitational wave}
\newacro{LIGO}[LIGO]{Laser Interferometer \acs{GW} Observatory}
\newacro{GCN}[GCN]{\acs{GRB} Coordinates Network}
\newacro{PTF}[PTF]{Palomar Transient Factory}
\newacro{iPTF}[iPTF]{intermediate Palomar Transient Factory}
\newacro{P48}[P48]{Palomar 48~inch Oschin telescope}
\newacro{P60}[P60]{Palomar 60~inch telescope}
\newacro{P200}[P200]{Palomar 200~inch telescope}
\newacro{OT}[OT]{optical transient}
\newacro{VLA}[VLA]{Karl G. Jansky Very Large Array}
\newacro{FOV}[FOV]{field of view}
\newacro{KAGRA}[KAGRA]{KAmioka GRAvitational\nobreakdashes-wave observatory}
\newacro{HETE}[HETE]{High Energy Transient Explorer}
\newacro{CBC}[CBC]{compact binary coalescence}
\newacro{NS}[NS]{neutron star}
\newacro{BH}[BH]{black hole}
\newacro{DBSP}[DBSP]{Double Spectrograph}
\newacro{ZTF}[ZTF]{Zwicky Transient Facility}
\newacro{NSF}[NSF]{National Science Foundation}
\newacro{FTN}[FTN]{Faulkes Telescope North}
\newacro{SED}[SED]{spectral energy distribution}
\newacro{BATSE}[BATSE]{Burst and Transient Source Experiment}
\newacro{PSF}[PSF]{point spread function}
\newacro{IPAC}[IPAC]{Infrared Processing and Analysis Center}
\newacro{CASA}[CASA]{Common Astronomy Software Applications}
\newacro{PC}[PC]{photon counting}
\newacro{UVOT}[UVOT]{Ultraviolet/Optical Telescope}
\newacro{BAT}[BAT]{Burst Alert Telescope}
\newacro{SMS}[SMS]{Short Message Service}
\newacro{LSB}[LSB]{long, soft burst}
\newacro{SHB}[SHB]{short, hard burst}
\newacro{SDSS}[SDSS]{Sloan Digital Sky Survey}
\newacro{CCD}[CCD]{charge coupled device}
\newacro{ToO}[ToO]{target of opportunity}
\newacroplural{ToO}[ToOs]{targets of opportunity}
\newacro{IMACS}[IMACS]{Inamori-Magellan Areal Camera \& Spectrograph}
\newacro{SNR}[S/N]{signal\nobreakdashes-to\nobreakdashes-noise ratio}
\newacro{HEALPix}[HEALPix]{Hierarchical Equal Area isoLatitude Pixelization}
\newacro{LSST}[LSST]{Large Synoptic Survey Telescope}

\begin{document}

\title{Discovery and redshift of an optical afterglow in 71 square degrees: \\
\lowercase{i}PTF13bxl and \acs{GRB}\,130702A}

\slugcomment{Received 2013 July 20; accepted 2013 September 9; published 2013 October 7}

\author{Leo P. Singer\altaffilmark{1}}
\author{S. Bradley Cenko\altaffilmark{2}}
\author{Mansi M. Kasliwal\altaffilmark{3,13}}
\author{Daniel A. Perley\altaffilmark{13,4}}
\author{Eran O. Ofek\altaffilmark{5}}
\author{Duncan A. Brown\altaffilmark{1,6}}
\author{Peter E. Nugent\altaffilmark{7,8}}
\author{S. R. Kulkarni\altaffilmark{4}}
\author{Alessandra Corsi\altaffilmark{9}}
\author{Dale A. Frail\altaffilmark{10}}
\author{Eric Bellm\altaffilmark{4}}
\author{John Mulchaey\altaffilmark{3}}
\author{Iair Arcavi\altaffilmark{5}}
\author{Tom Barlow\altaffilmark{4}}
\author{Joshua S. Bloom\altaffilmark{7,8}}
\author{Yi Cao\altaffilmark{4}}
\author{Neil Gehrels\altaffilmark{2}}
\author{Assaf Horesh\altaffilmark{4}}
\author{Frank J. Masci\altaffilmark{11}}
\author{Julie McEnery\altaffilmark{2}}
\author{Arne Rau\altaffilmark{12}}
\author{Jason A. Surace\altaffilmark{11}}
\author{Ofer Yaron\altaffilmark{5}}

\email{lsinger@caltech.edu}

\altaffiltext{1}{LIGO Laboratory, California Institute of Technology, Pasadena, CA 91125, USA}
\altaffiltext{2}{Astrophysics Science Division, NASA Goddard Space Flight Center, Mail Code 661, Greenbelt, MD 20771, USA}
\altaffiltext{3}{Observatories of the Carnegie Institution for Science, 813 Santa Barbara St, Pasadena CA 91101, USA}
\altaffiltext{4}{Cahill Center for Astrophysics, California Institute of Technology, Pasadena, CA 91125, USA}
\altaffiltext{5}{Benoziyo Center for Astrophysics, The Weizmann Institute of Science, Rehovot 76100, Israel}
\altaffiltext{6}{Department of Physics, Syracuse University, Syracuse, NY 13244, USA}
\altaffiltext{7}{Department of Astronomy, University of California Berkeley, B-20 Hearst Field Annex \# 3411, Berkeley, CA, 94720-3411}
\altaffiltext{8}{Physics Division, Lawrence Berkeley National Laboratory, 1 Cyclotron Road MS 50B-4206, Berkeley, CA 94720, USA}
\altaffiltext{9}{George Washington University, Corcoran Hall, Washington, DC 20052, USA}
\altaffiltext{10}{National Radio Astronomy Observatory, P.O. Box O, Socorro, NM 87801, USA}
\altaffiltext{11}{Infrared Processing and Analysis Center, California Institute of Technology, Pasadena, CA 91125, USA}
\altaffiltext{12}{Max-Planck-Institut f\u{u}r extraterrestrische Physik, Giessenbachstrasse 1, 85748 Garching, Germany}
\altaffiltext{13}{Hubble Fellow}

\shortauthors{Singer et al.}


%TC:break Abstract
\begin{abstract}
We report the discovery of the optical afterglow of the \ac{GRB}\,130702A, identified upon searching 71\,deg$^2$ surrounding the \emph{Fermi} \ac{GBM} localization. Discovered and characterized by the \acl{iPTF}, iPTF13bxl is the first afterglow discovered solely based on a \ac{GBM} localization. Real\nobreakdashes-time image subtraction, machine learning, human vetting, and rapid response multi\nobreakdashes-wavelength follow\nobreakdashes-up enabled us to quickly narrow a list of 27,004 optical transient candidates to a single afterglow-like source. Detection of a new, fading X\nobreakdashes-ray source by \emph{Swift} and a radio counterpart by \acs{CARMA} and the \acl{VLA} confirmed the association between iPTF13bxl and \ac{GRB}\,130702A. Spectroscopy with the Magellan and  Palomar 200 inch telescopes showed the afterglow to be at a redshift of $z=0.145$, placing \ac{GRB}\,130702A among the lowest redshift \acp{GRB} detected to date.  The prompt $\gamma$\nobreakdashes-ray energy release and afterglow luminosity are intermediate between typical cosmological \acp{GRB} and nearby sub-luminous events such as GRB\,980425 and GRB\,060218.  The bright afterglow and emerging supernova offer an opportunity for extensive panchromatic follow\nobreakdashes-up. Our discovery of iPTF13bxl demonstrates the first observational proof\nobreakdashes-of\nobreakdashes-principle for $\sim$10 \emph{Fermi}\nobreakdashes-iPTF localizations annually.  Furthermore, it represents an important step toward overcoming the challenges inherent in uncovering faint optical counterparts to comparably localized gravitational wave events in the Advanced \acs{LIGO} and Virgo era.
\end{abstract}
%TC:break _main_

\section{Introduction}

%Compton/BATSE 20-600 keV
%BeppoSAX - 40-700 KeV - all sky -  the first afterglow - the first redshift -
%Swift/BAT - 15-150 keV - XRT+BAT - large N game - 0.5 sr, 4% of sky
%Fermi/GBM - few keV to 1 MeV, 150 keV to 30 MeV - 8.8 sr, 70% of sky
%Fermi/LAT -  20 MeV to 300 GeV - >2sr, 16% of sky

Our understanding of \acfp{GRB} has been propelled by our ability to localize these rare and energetic cosmic events precisely. \emph{Compton Gamma\nobreakdashes-ray Observatory}/BATSE's coarse localizations robustly demonstrated that \acp{GRB} were distributed isotropically on the sky and suggested that \acp{GRB} originate at cosmological distances~\citep{GRBsAreExtragalactic}. Prompt arcminute localizations provided by \emph{BeppoSAX} directly enabled the discovery of the first afterglows of long-duration \acp{GRB} \citep{GRBsHaveXrayAfterglows,GRBsHaveOpticalAfterglows,GRBsHaveRadioAfterglows}.  
Currently, the prompt slewing capabilities of the \textit{Swift} satellite 
\citep{gcg+04} enable the on\nobreakdashes-board narrow-field instruments to provide arcsecond localizations for $\approx 90$ \acp{GRB} yr$^{-1}$ within $\approx 100$\,s of the burst trigger.

With seven decades of simultaneous energy coverage, \emph{Fermi} has opened a new window into the \ac{GRB} phenomenon, the MeV to GeV regime.  However, \emph{Fermi} remains fundamentally limited by its localization capabilities.  The \aclu{LAT} \citep[\acs{LAT}; 20\,MeV\nobreakdashes--300\,GeV; 16\% of all-sky;][]{LAT} can localize events with GeV photons to radii as small as $\sim 10$\arcmin.  But the \ac{LAT} only localizes a handful of \acp{GRB} each year.  The \aclu{GBM} (\acs{GBM}; few\,keV--30\,MeV; 70\% of all-sky; \citealt{FermiGBM}), on the other hand, detects \acp{GRB} at a rate of $\approx 250$\,yr$^{-1}$.  However, typical \ac{GBM} \acp{GRB} have localizations of many tens of square degrees (random plus systematic uncertainties).  Consequently, no afterglows have been identified based solely on a \ac{GBM} localization until this work\footnote{The only comparable discovery was the afterglow of GRB\,120716A in the $\approx 2$\,deg$^{2}$ error box from the \ac{IPN} by \citet{GCN13489}.}.

\begin{figure*}
    \includegraphics{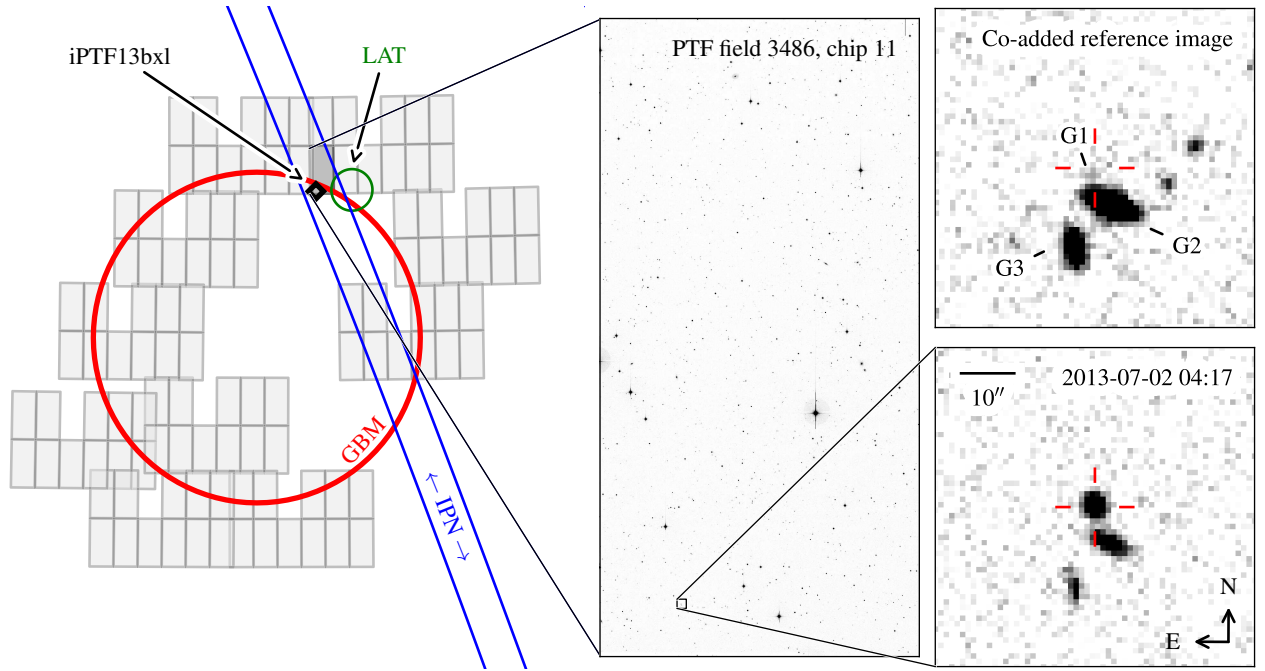}
    \caption{\label{fig:discovery}P48 imaging of GRB\,130702A and discovery of iPTF13bxl. The left panel illustrates the $\gamma$-ray localizations (red circle: 1$\sigma$ \ac{GBM}; green circle: \ac{LAT}; blue lines: 3$\sigma$ \ac{IPN}) and the 10 \ac{P48} reference fields that were imaged (light gray rectangles).  For each P48 pointing, the location of the 11 chips are indicated with smaller rectangles (one CCD in the camera is not currently operable).  Our tiling algorithm places a large weight on the existence of deep P48 pre-explosion imaging (a necessity for high-quality subtraction images); the large gaps inside the GBM localization are fields without these reference images.  The small black diamond is the location of iPTF13bxl.  The right panels show \ac{P48} images of the location of iPTF13bxl, both prior to (top) and immediately following (bottom) discovery.  We note that the \ac{LAT} and \ac{IPN} localizations were published \textit{after} our discovery announcement \citep{GCN14967}. \\ (A color version of this figure is available in the online journal.)}
\end{figure*}

The \acl{PTF} (\acsu{PTF}; \citealt{PTF}) is developing the necessary instrumentation, algorithms, and observational discipline to detect optical counterparts to \ac{GBM} \acp{GRB}. The wide 7.1\,deg$^2$ \acl{FOV} and sensitivity ($R \approx 20.6$\,mag in 60\,s) of the \ac{P48} and CFHT12k camera~\citep{rsv+08} are well-suited to identifying long-duration \ac{GRB} afterglow candidates. The real\nobreakdashes-time software pipeline (Nugent et al., in prep.) enables rapid panchromatic follow-up with an arsenal of telescopes~(e.g. \citealt{2011ApJ...736..159G}), essential to distinguish the true afterglow from background and foreground contaminants.  Here, we present our discovery of iPTF13bxl, the afterglow of the \emph{Fermi}~\ac{GBM} \ac{GRB}~130702A found by searching a sky area of 71\,deg$^2$ with the \acl{iPTF} (\acsu{iPTF}).

\section{Discovery}
\label{sec:discovery}

On 2013~July~2 at 00:05:23.079~UT, the \emph{Fermi} \ac{GBM} detected trigger 394416326.  The refined human-generated (i.e., ground\nobreakdashes-based) localization, centered on $\alpha = 14^{\mathrm{h}} 35^{\mathrm{m}} 14^{\mathrm{s}}$, $\delta = 12^{\circ} 15\arcmin 00\arcsec$ (J2000.0), with a quoted 68\% containment radius of $4\fdg0$ (statistical uncertainty only), was disseminated less than an hour after the burst (Figure~\ref{fig:discovery}).

\begin{figure*}
   \includegraphics{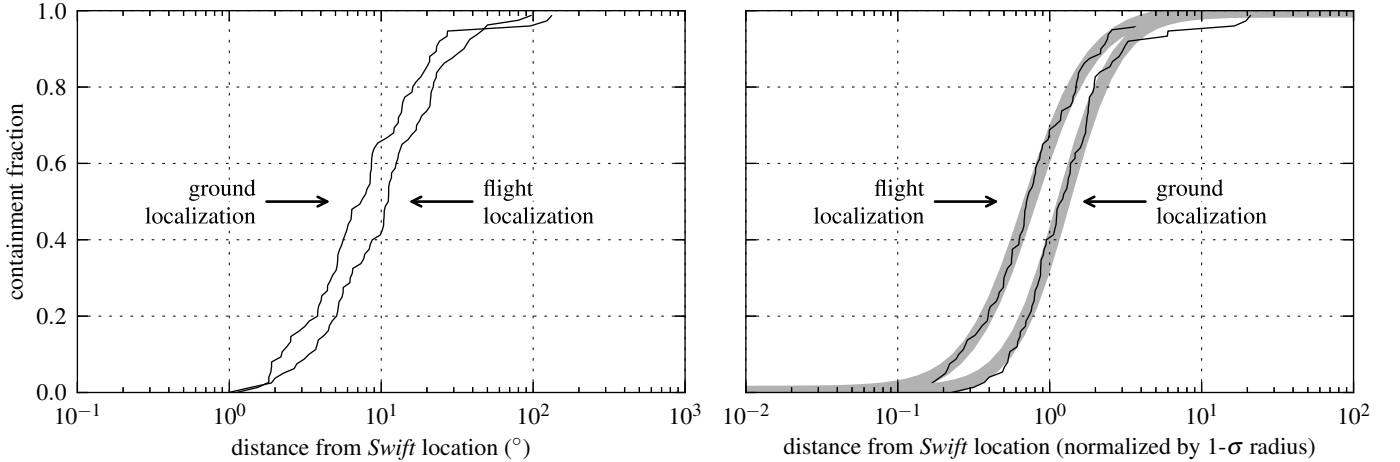}
   \caption{\label{fig:fermi-localization}Localization accuracy of \emph{Fermi} \ac{GBM} positions, generated by searching for coincidences with \acp{GRB} detected by the \emph{Swift} satellite.  The left panel shows the fraction of bursts contained within a given distance from the \emph{Swift} position, both for in\nobreakdashes-flight and refined ground\nobreakdashes-based localizations.  Ground-based localizations are on average about half as far from the true \ac{GRB} positions as the in-flight localizations. The right panel shows a cumulative histogram of the \emph{Fermi}\nobreakdashes--\emph{Swift} distance, normalized by each trigger's nominal 1$\sigma$ radius (either ground-based or in-flight).  Although the ground-based localizations are more accurate, the nominal ground\nobreakdashes-based error radii must be interpreted as describing a different confidence level than the  in\nobreakdashes-flight error radius. The thick gray lines are fits to the logistic\nobreakdashes-like function in Equation~\ref{eq:fermi-localization}.}
\end{figure*}

\emph{Fermi}-\ac{GBM} \ac{GRB} positions are known to suffer from significant systematic uncertainties, currently estimated to be $\approx 2^\circ$--$3^\circ$.  To characterize the full radial profile of the localization uncertainty, our \ac{GBM}\nobreakdashes-\ac{iPTF} pipeline automatically computed a probability map for the event modeled on previous \emph{Fermi}/\emph{Swift} coincidences from 2010~March~30 through 2013~April~4. We fit a sigmoid function:
\begin{equation}
    \label{eq:fermi-localization}
    P(r) = \frac{1}{1 + \left(c_0 r\right)^{c_1}}
\end{equation}
where $r$ is the angular distance to the \emph{Swift} location, normalized by the in\nobreakdashes-flight or ground\nobreakdashes-based error radius for that burst. We find $c_0 = 1.35$, $c_1 = -2.11$ for in\nobreakdashes-flight \ac{GBM} localizations and $c_0 = 0.81$, $c_1 = -2.47$ for ground\nobreakdashes-based \ac{GBM} localizations (Figure~\ref{fig:fermi-localization}). 

Image subtraction within \ac{iPTF} is greatly simplified by observing only pre-defined fields on the sky; this ensures that sources will fall on approximately the same detector location from night to night, minimizing a possible source of systematic uncertainty.  Using a \acl{HEALPix}~\citep[\acsu{HEALPix};][]{HEALPix} bitmap representation of the probability distribution of the trigger location, we chose 10 of these pre-defined fields to maximize the probability of enclosing the true (but still unknown) location of the burst (Figure~\ref{fig:discovery}). In this particular case, the ten selected fields did not include the center of the \ac{GBM} localization because we lacked previous reference images there. Nonetheless, we estimated that these ten fields had a 38\% chance of containing this \ac{GRB}'s location. Given the youth, sky location, and probability of containment, we let our software trigger follow\nobreakdashes-up with the \ac{P48}.

Starting at 04:17:23~UT ($\Delta t \equiv t - t_{\mathrm{GBM}} = 4.2$\,hr), we imaged each of these ten fields twice in 60\,s exposures with the Mould $R$ filter. These fields were then subjected to the standard \ac{iPTF} transient search: image subtraction, source detection, and ``real/bogus'' machine ranking \citep{BloomMachineLearning,brp+12} into likely astrophysical transient sources (``real'', or 1) or likely artifacts (``bogus'', or 0).

The \ac{iPTF} real\nobreakdashes-time analysis found 27,004 transient/variable candidates in these twenty individual subtracted images. Of these, 44 were eliminated because they were determined to be known asteroids in the Minor Planet Checker database\footnote{\url{http://www.minorplanetcenter.org/iau/mpc.html}} using PyMPChecker.\footnote{\url{http://dotastro.org/PyMPC/PyMPC/}} Demanding a real/bogus rank greater than 0.1 reduced the list to 4214.  Rejecting candidates that coincided with point sources in \ac{SDSS} brighter than $r'=21$ narrowed this to 2470.  Further requiring detection in both \ac{P48} visits and imposing \acs{CCD}\nobreakdashes-wide data quality cuts (e.g., bad pixels) eliminated all but 43 candidates.  Following human inspection, seven sources were saved as promising transients in the \ac{iPTF} database.

Two candidates, iPTF13bxh and iPTF13bxu, were near the cores of bright, nearby galaxies, environments that are inherently variable and also present a challenge to image subtraction. A third, iPTF13bxr, was coincident with a galaxy in \ac{SDSS} with a quasar spectrum (SDSS\,J145359.72+091543.3). iPTF13bxt was close to a star in \ac{SDSS}, and so was deemed a likely variable star.  We did not consider these further for the afterglow search. The final three sources, iPTF13bxj (real-bogus score $= 0.86$), iPTF13bxk (real-bogus score $= 0.49$), and iPTF13bxl (real-bogus score $= 0.83$), remained as potential counterparts and were scheduled for $g'r'i'$ photometry with the \acl{P60}~\citep[\acsu{P60};][]{P60Automation} and spectroscopic classification on the \ac{P200}.  iPTF13bxl ($\alpha = 14^\mathrm{h}29^\mathrm{m}14\fs78, \delta = +15\arcdeg46\arcmin, 26\farcs4$) was immediately identified as the most promising candidate because it showed a significant intra\nobreakdashes-night decline. Our panchromatic follow-up (Sections~\ref{sec:followup}~and~\ref{sec:spec})
confirmed iPTF13bxl was indeed the afterglow. Subsequent spectroscopy revealed iPTF13bxj to be a Type II supernova at $z=0.06$ and iPTF13bxk a quasar at $z=2.4$.

Following our discovery announcement \citep{GCN14967}, the \emph{Fermi} \ac{LAT} and \ac{GBM} teams published \ac{GCN} circulars announcing the detection of \ac{GRB}~130702A~\citep{GCN14971,GCN14972}.  As seen by the \ac{GBM}, GRB\,130702A had a duration of $t_{90} \approx 59$\,s and a 10\,keV\nobreakdashes--1\,MeV fluence of $f_{\gamma} = (6.3 \pm 2.0) \times 10^{-6}$\,erg\,cm$^{-2}$.  The best-fit power-law spectrum may suggest a classification as an X-ray flash.  The \ac{LAT} location was $0\fdg9$ from iPTF13bxl, with a 90\% statistical error radius of $0\fdg5$. An \ac{IPN} triangulation \citep{GCN14974} yielded a 3\nobreakdashes-$\sigma$ annulus that was $0\fdg46$ wide from its center to its edges. iPTF13bxl was $0\fdg16$ from the annulus' centerline (Figure~\ref{fig:discovery}).

\section{Broadband photometric follow\nobreakdashes-up}
\label{sec:followup}

On 2013~July~3 at 4:10~UT, ($\Delta t = 28.1$\,hr), the \ac{P60} obtained two sequences of Sloan $g'r'i'$ observations of the field of iPTF13bxl. \ac{P60} observations were calibrated relative to 20 reference stars in the \ac{SDSS} (AB) system. Final reduction of the \ac{P48} observations was performed automatically at the \acl{IPAC}~\citep{P48PhotometricCalibration}. We corrected the \ac{P48} and \ac{P60} photometry for Galactic extinction using maps from \citet[][$E(B-V) = 0.024$\,mag]{SchlaflyExtinction}.

\begin{figure*}
    \includegraphics{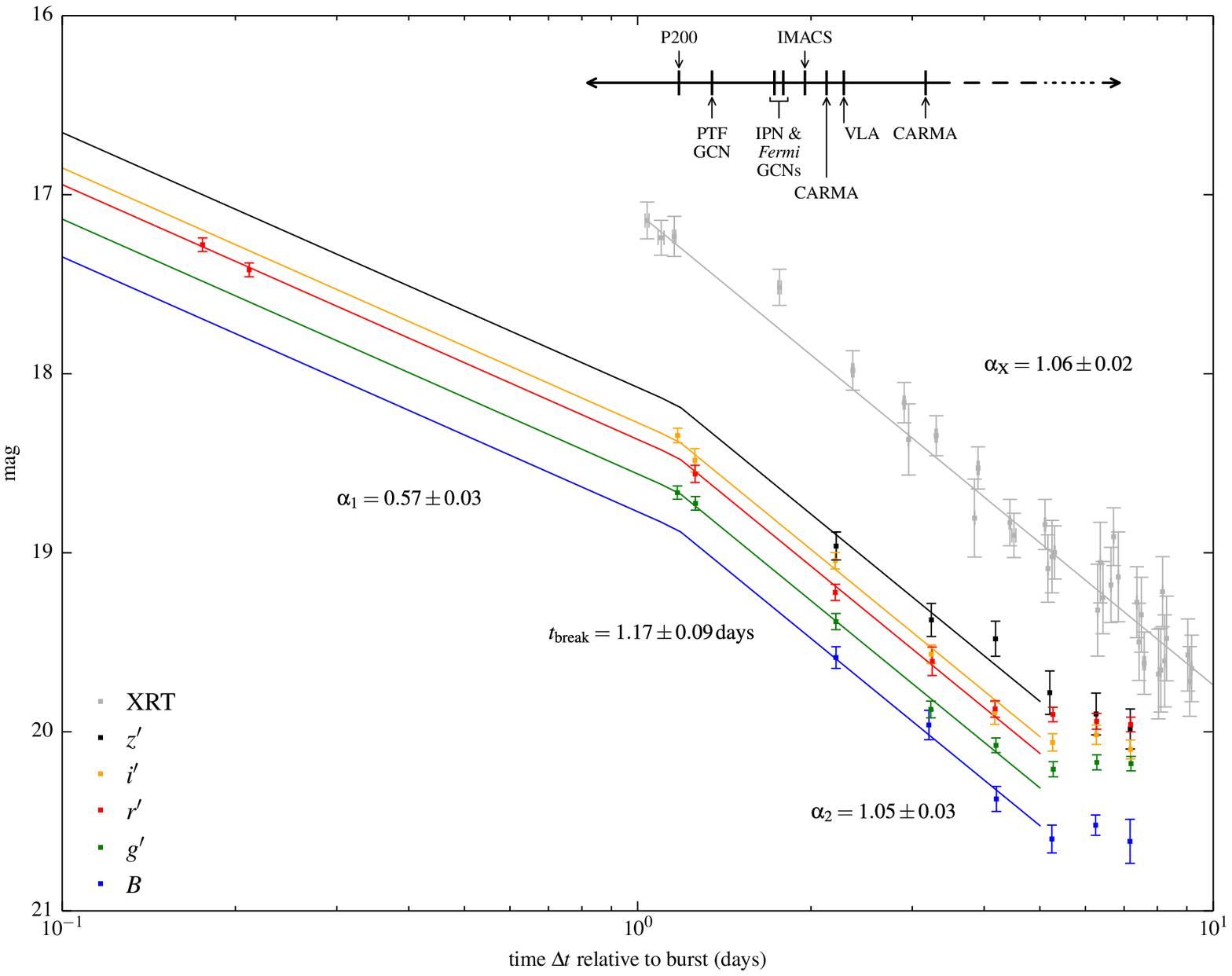}
    \caption{\label{fig:lightcurve}\ac{P48}, \ac{P60}, and \ac{XRT} light curves of iPTF13bxl.  The broken power\nobreakdashes-law fit is shown up to $\Delta t=5$\,days. The \ac{XRT} observations, re-binned to improve presentation, are shown in gray as $m(\mathrm{AB})-6.5$ at 1\,keV. A timeline in the top right puts the \ac{P48} and \ac{P60} observations in the context of the \ac{XRT} follow\nobreakdashes-up, \ac{PTF}'s discovery \ac{GCN}~\citep{GCN14967}, the announcement of the \ac{LAT}~\citep{GCN14971} and \ac{IPN}~\citep{GCN14974} localizations, and the radio observations. \\ (A color version of this figure is available in the online journal.)}
\end{figure*}

\begin{figure*}
    \includegraphics{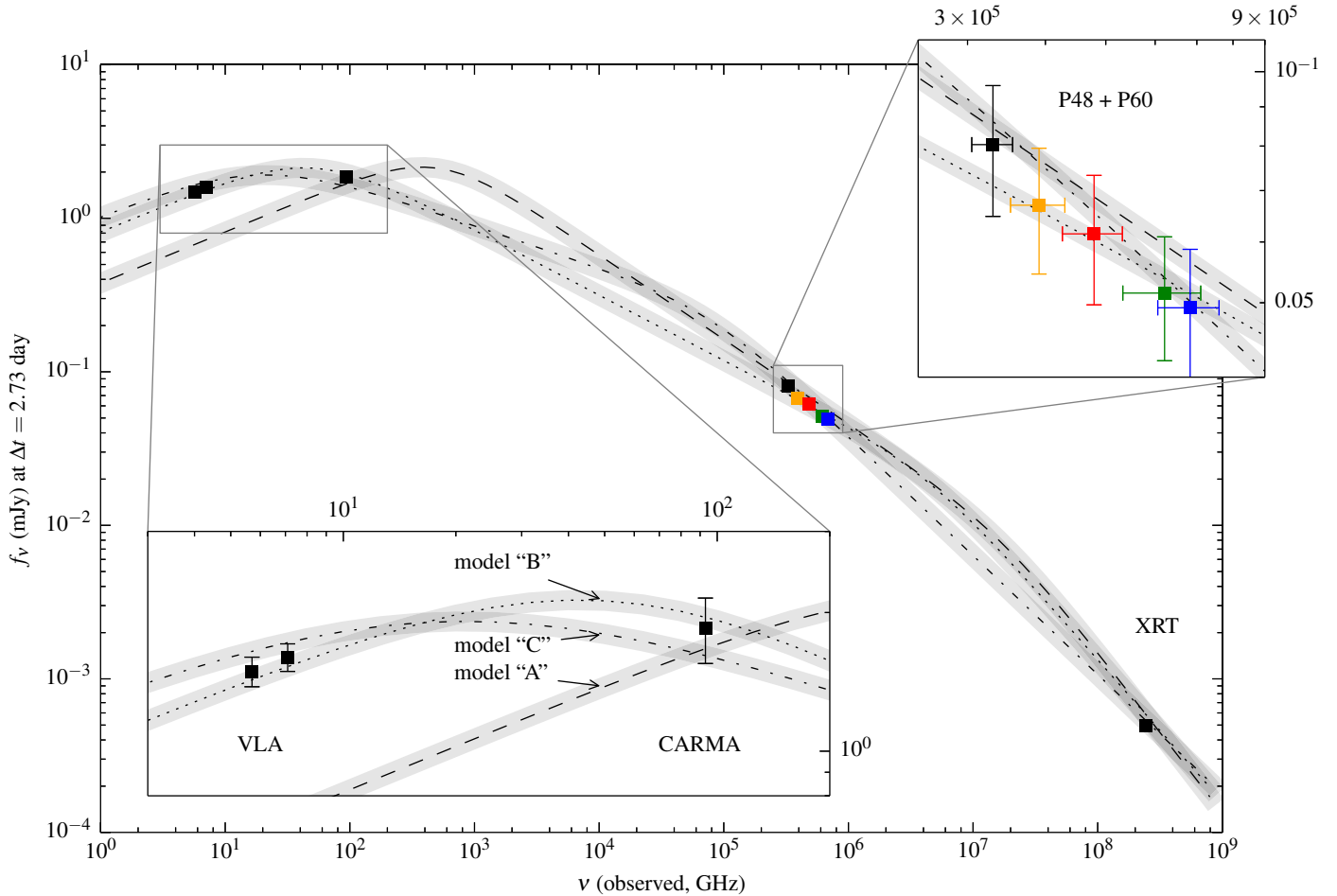}
    \caption{\label{fig:sed}Broadband \ac{SED} of iPTF13bxl. Two insets show details of the radio and optical observations respectively. The \ac{XRT} and optical observations have been interpolated to the mean time of the radio observations ($\Delta t = 2.6$\,days). \\ (A color version of this figure is available in the online journal.)}
\end{figure*}

The optical light curve is shown in Figure~\ref{fig:lightcurve}. We fit an achromatic broken power law to all filters and all times up to $\Delta t=5$\,days after the burst. A spectral slope of $\beta_\mathrm{O}=0.7 \pm 0.1$ is sufficient to characterize the post\nobreakdashes-break color, illustrated in the inset of Figure~\ref{fig:sed}. We note that the optical decay ceased at $r' \approx 20$\,mag after $\Delta t \approx\,5$\,days when the accompanying supernova started to dominate \citep{GCN14994}. This supernova will be the subject of a future work.

Following our discovery of iPTF13bxl, we triggered \acl{ToO} observations with the \emph{Swift} \ac{XRT} \citep{bhn+05} beginning at 00:50~UT on 2013~July~3 ($\Delta t = 1.03$\,days).  We downloaded the data products from the \emph{Swift} \ac{XRT} repository \citep{ebp+07}. The resulting 0.3\nobreakdashes--10\,keV light curve is plotted in Figure~\ref{fig:lightcurve}. The spectrum is well fit by a power law with photon index $\Gamma = 2.0^{+0.14}_{-0.13}$, while the light curve fades in time with a power-law slope of $\alpha_\mathrm{X} = 1.06 \pm 0.02$, in excellent agreement with the post\nobreakdashes-break optical decay.

After the discovery of the optical counterpart to \ac{GRB}\,130702A, we began observations with the \ac{CARMA}. All observations were carried out in single\nobreakdashes-polarization mode with the 3\,mm receivers tuned to a frequency of 93\,GHz, and were reduced using MIRIAD. We flux\nobreakdashes-calibrated the data using observations of MWC349 and 3C273.  The afterglow is well-detected in both epochs, and we measure flux densities of $1.58 \pm 0.33$\,mJy and $1.85 \pm 0.30$\,mJy on July 4.13 and 5.17, respectively.

The position of iPTF13bxl was observed with the \ac{VLA} in C\nobreakdashes-band beginning at 6:20~UT on 2013~July~4 ($\Delta t = 2.29$\,days).  The observations were conducted using the standard WIDAR correlator setting. Data were reduced using the Astronomical Image Processing System package following standard practice. 3C286 was used for bandpass and flux calibration; J1415+1320 was used for gain and phase calibration. We detect a radio source with flux density of $1.49 \pm 0.08$\,mJy at 5.1\,GHz at $1.60 \pm 0.08$\,mJy at 7.1\,GHz. Errors on the measured flux were calculated as the quadrature sum of the map root\nobreakdashes-mean square and a fractional systematic error (of the order of 5\%) to account for uncertainty in the flux density calibration.

The broadband \ac{SED} is shown in Figure~\ref{fig:sed}. We interpolated both the optical and X-ray observations to the mean time of the \ac{VLA} and \ac{CARMA} observations.  In the context of the standard synchrotron afterglow model~\citep{SariPiranNarayan98}, the comparable X-ray and optical spectral and temporal indices at this time suggest both bandpasses lie in the same spectral regime, likely with $\nu > \nu_{c}$.  This would imply a relatively hard electron spectral energy index ($N(\gamma_e) \propto \gamma_e^{-p}$) $p \approx 2$, possibly requiring a high-energy cut-off.  

Also in Figure~\ref{fig:sed} we plot three broadband \ac{SED} models synthesized using techniques similar to \citet{pcc+13}. Although these models are not formal fits to our highly under-constrained observations, they demonstrate overall consistency with standard synchrotron afterglow theory.  Model ``A'' (dashed line; $\chi^2 = 126$) represents a constant\nobreakdashes-density (ISM) circumburst medium with $p = 2.1$, $\epsilon_B = 0.48$, $\epsilon_e = 0.41$, $E = 3\times10^{51}$\,erg, $n=1.2 \times 10^{-3}$\,cm$^{-3}$.  This model under-predicts the \ac{VLA} bands, but this deviation could be due to scintillation or reverse shock emission at low frequencies.  Model ``B'' (dotted line; $\chi^2 = 7$) is in a wind environment ($\rho \propto r^{-2}$) with $p = 2.1$, $\epsilon_B = 0.32$, $\epsilon_e = 0.43$, $E = 1.4\times10^{51}$\,erg, $A* = 4.8\times10^{-3}$\,g\,cm$^{-1}$.  This fits the data well except for a small discrepancy with the optical spectral slope.  Model ``C'' (dotted-dashed line; $\chi^2 = 6$) is a similar wind model but with $p = 1.55$.  Of the three, ``C'' fits the data best, but seems non\nobreakdashes-physical (high\nobreakdashes-energy cutoff required).  Accurate determination of the underlying physical parameters would require tracing the evolution of the SED with time. 

\section{Optical Spectroscopy and Host Galaxy Environment}
\label{sec:spec}

\begin{figure*}
    \includegraphics{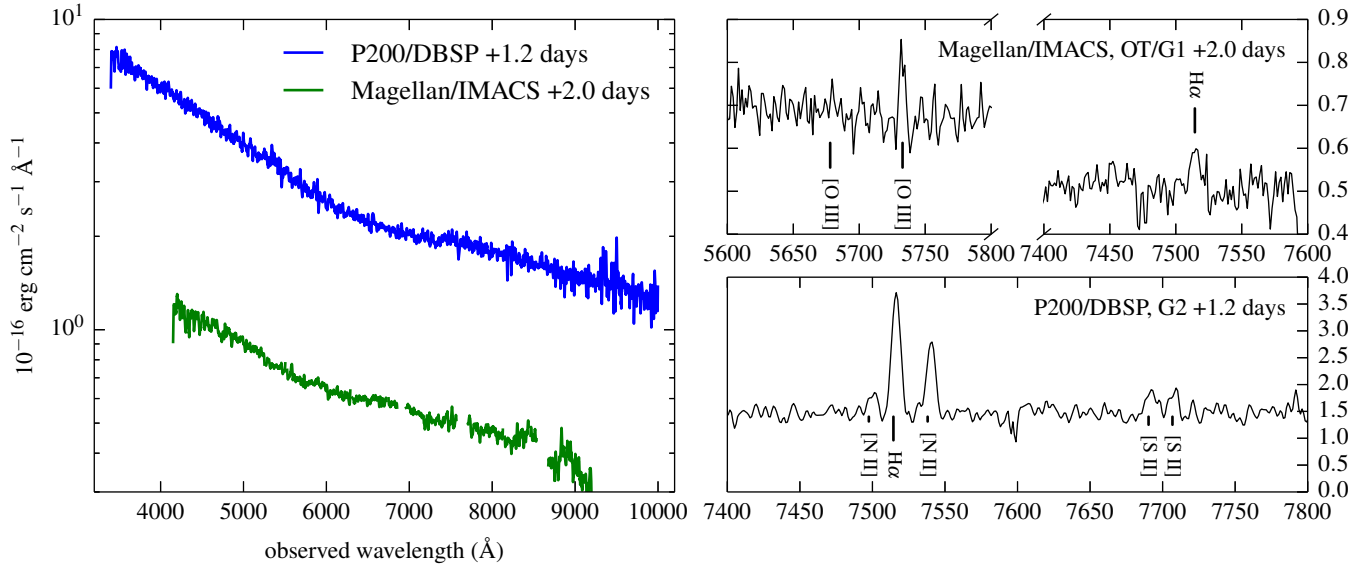}
    \caption{\label{fig:spectra}Optical spectra of iPTF13bxl and the nearby galaxy SDSS\,J142914.57+154619.3 (``G2''). Spectra in the left panel have been smoothed with a Savitzky-Golay filter. Our initial P200 spectrum of the afterglow (left panel, blue) exhibits a largely featureless blue continuum.  A higher \ac{SNR} spectrum taken the following night with \ac{IMACS} (left panel, green) revealed faint emission features corresponding to [\ion{O}{3}] and H$\alpha$ at $z = 0.145$ (top right panel). The bottom right panel shows a spectrum of the nearby galaxy G2, which has the same redshift as iPTF13bxl. \\ (A color version of this figure is available in the online journal.)}
\end{figure*}

We obtained optical spectra of iPTF13bxl with the \ac{DBSP} mounted on the \ac{P200} on 2013~July~3.17 and the \acl{IMACS}~\citep[\acsu{IMACS};][]{IMACS} mounted on the 6\,m Magellan Ba'ade telescope on 2013~July~3.97 ($\Delta t = 1.2$ and 2.0\,days, respectively).  The resulting spectra are plotted in Figure~\ref{fig:spectra}.

Our initial \ac{DBSP} spectrum exhibits a largely featureless, blue continuum. The higher \ac{SNR} of the \ac{IMACS} spectrum further reveals faint, narrow emission lines corresponding to [\ion{O}{3}] and H$\alpha$ at a common redshift of $z = 0.145 \pm 0.001$ (luminosity distance $d_{L} = 680$\,Mpc), which we adopt as the distance to GRB\,130702A. The continuum of both spectra exhibit deviations from a single power-law, with excess flux (when compared with the late-time photometric spectral index of $\beta_\mathrm{O} = 0.7$) visible at shorter wavelengths.  This may suggest some contribution from either shock breakout or the emerging supernova at very early times post-explosion.

Three galaxies are visible in the immediate environment of iPTF13bxl in pre-explosion imaging (labeled ``G1'', ``G2'', and ``G3'' in Figure~\ref{fig:discovery}).  Presumably the emission lines observed from iPTF13bxl arise in G1, given the small spatial offset ($0\farcs6$) and slit orientation (PA $= 90$).  However, our spectra also reveal galaxies G2 and G3 both lie at redshifts consistent with iPTF13bxl (e.g., Figure~\ref{fig:spectra}).  Observations with DEIMOS on the Keck~II telescope reveal two more galaxies at the same redshift at separations of $1\farcm2$ (SDSS~J142910.29+154552.2) and $2\farcm7$ (SDSS~J142917.67+154352.2) from the transient. The explanation most consistent with past observations of long-duration \ac{GRB} host galaxies (e.g., \citealt{sgl09}) is that GRB\,130702A exploded in a dwarf ($M_{r} \approx -16$\,mag) member of this association or group, a relatively unusual environment \citep[see also][]{HostEnvironment}.

\section{\ac{GRB}\,130702A in Context}
\label{sec:context}
The prompt $\gamma$-ray isotropic energy release ($E_{\gamma,\mathrm{iso}}$) of \acp{GRB} spans a range of six orders of magnitude, from $\sim 10^{48}$--$10^{54}$\,erg.  At $z = 0.145$, the prompt emission from \ac{GRB}\,130702A constrains $E_{\gamma,\mathrm{iso}} \lesssim (6.5 \pm 0.1) \times 10^{50}$\,erg~\citep[90\% upper limit;][]{GCN15025}. This value is significantly smaller than typical cosmological \acp{GRB} ($E_{\gamma,\mathrm{iso}} \sim 10^{52}$--$10^{54}$\,erg; \citealt{a06,bkb+07}).  Yet GRB\,130702A greatly outshines the most nearby, sub-luminous events with well-studied supernovae, such as GRB\,980425 ($E_{\gamma,\mathrm{iso}} = 1.0 \times 10^{48}$\,erg; \citealt{paa+00}) and GRB\,060218 ($E_{\gamma,\mathrm{iso}} = 6.2 \times 10^{49}$\,erg; \citealt{cmb+06}).  

At all wavelengths, the counterpart behaves like a typical \ac{GRB} afterglow scaled down in luminosity by a factor of $\sim$10 compared to a ``typical'' \emph{Swift} burst (or $\sim$100 lower than a luminous pre-\emph{Swift} burst) as observed at the same time.  This is intuitively explained by an equivalent scaling down of the overall energy (per solid angle) of the burst and shockwave relative to more typical, high-luminosity bursts.  It is not yet clear whether this energy difference is due primarily to the release of less relativistic ejecta by the burst overall, a wider jet, or a partially off-axis view of a structured jet.  Late-time radio follow-up should help distinguish these models: an intrinsically low-energy \ac{GRB} should produce a much earlier jet break than a widely-beamed burst, while a structured jet will actually produce an \emph{increase} in flux at late times as the jet core spreads and its radiation enters our sightline.

Events with similar energetics have been found by \emph{Swift}, e.g., \ac{GRB}\,050826 at $z=0.30$ and \ac{GRB}\,120422A at $z=0.28$ \citep{2007ApJ...661L.127M,zfs+12}.  However, given their low intrinsic luminosities and higher redshift, the afterglows were too faint to identify late-time breaks and establish their shock energies $E_K$, making them difficult to physically interpret.  \ac{GRB}\,130702A's proximity avoids both these problems. Our observations suggest---and further observations should confirm---that its $\gamma$-ray and afterglow energetics are intermediate between these two previously quite-disparate classes of \acp{GRB}, helping to fill in the ``gap'' between the well-studied cosmological population and the class of less-luminous local \acp{GRB} and relativistic Type Ic supernovae (e.g., \citealt{2004Natur.430..648S,scp+10}).

\section{Conclusion}
\label{sec:conclusion}
Using the infrastructure outlined above, we estimate that a dedicated \ac{iPTF} program would recover $\sim$10 \ac{GBM} afterglows each year.  The addition of other surveys with comparably wide \aclp{FOV} and apertures (e.g., Pan\nobreakdashes-STARRS, SkyMapper, CRTS) could increase this number, assuming they had similar real\nobreakdashes-time transient detection and follow-up programs in place.  Since \ac{GBM} detects \acp{GRB} in the classical $\gamma$-ray band, their optical counterparts should more closely resemble the pre-\textit{Swift} population ($\approx 1$\,mag brighter at a fixed time; \citealt{kkz+10}).  Even if only a single event per year as nearby as \ac{GRB}\,130702A were uncovered, this would still represent a remarkable advance in our understanding of the \ac{GRB}\nobreakdashes--supernova connection.

Furthermore, this work sets the stage for more discoveries in ongoing and future physics experiments that are limited by similarly coarse position reconstruction. Later this decade, a network of advanced \ac{GW} detectors including the \acl{LIGO} (\acsu{LIGO}) and Virgo is expected to detect $\sim 0.4$\nobreakdashes--400 binary neutron star mergers per year~\citep{LIGORates}, but with positions uncertain to tens to hundreds of deg$^2$~\citep{FairhurstTriangulation,NissankeLocalization,LIGOObservingScenarios}.

Optical counterparts to \ac{GW} sources will rarely (due to jet collimation) include bright, on\nobreakdashes-axis short\nobreakdashes-hard burst afterglows. Fainter $r$\nobreakdashes-process\nobreakdashes--fueled kilonovae \citep{kilonova} or yet fainter off\nobreakdashes-axis afterglows \citep{offaxis} are expected to accompany binary neutron star mergers. Both of these signatures are predicted to be several magnitudes fainter than iPTF13bxl. Optical searches will be inundated with astrophysical false positives~\citep{NissankeKasliwalEMCounterparts}. This problem will only be exacerbated for future surveys covering larger areas (e.g., \acl{ZTF}; \citealt{ZTF}) and/or with larger apertures (e.g., \acl{LSST}; \citealt{LSST}).  However, a breathtakingly complete astrophysical picture could reward us: masses and spins measured in \acp{GW}; host galaxy and disruption ejecta in optical; circumstellar environment in radio. The case of \ac{GRB}\,130702A demonstrates for the first time that optical transients can be recovered from localization areas of $\sim$100\,deg$^2$, reaching a crucial milestone on the road to Advanced \ac{LIGO}.

\acknowledgements Optical photometry and spectroscopy referred to in this work will be made available via WISeREP\footnote{\url{http://www.weizmann.ac.il/astrophysics/wiserep/}} \citep{yg12}.

We acknowledge A. Weinstein, A. Gal-Yam, R. Quimby, V. Connaughton, and the \emph{Fermi}-\ac{GBM} team for valuable discussions, S. Caudill, S. Tinyanont, D. Khatami for \ac{P200} observing, and the developers of the COSMOS package for Magellan data reduction.

This research is supported by the \acl{NSF} through a Graduate Research Fellowship for L.P.S., award PHY-0847611 for D.A.B., and NSF-CDI grant 0941742 for J.S.B.  M.M.K. acknowledges generous support from the Carnegie-Princeton Fellowship. M.M.K. and D.A.P. are supported by NASA through the Hubble Fellowship grants HST\nobreakdashes-HF\nobreakdashes-51293.01 and HST\nobreakdashes-HF\nobreakdashes-51296.01\nobreakdashes-A, awarded by the Space Telescope Science Institute, which is operated by the Association of Universities for Research in Astronomy, Inc., for NASA, under contract NAS~5\nobreakdashes-26555. E.O.O. is the incumbent of
the Arye Dissentshik career development chair and is supported by grants from the Israeli Ministry of Science and the I-CORE Program.  D.A.B. is further supported by an RCSA Cottrell Scholar award.

This research made use of Astropy \citep[][\url{http://www.astropy.org}]{astropy}, a community-developed core Python package for Astronomy.  The National Radio Astronomy Observatory is a facility of the \acl{NSF} operated under cooperative agreement by Associated Universities, Inc.

\bibliographystyle{apj}
\bibliography{ms}

\begin{thebibliography}{50}
\expandafter\ifx\csname natexlab\endcsname\relax\def\natexlab#1{#1}\fi

\bibitem[{{Aasi} {et~al.}(2013){Aasi}, {Abadie}, {Abbott}, {Abbott}, {Abbott},
  {Abernathy}, {Accadia}, {Acernese}, \& et~al.}]{LIGOObservingScenarios}
{Aasi}, J., {et~al.} 2013

\bibitem[{{Abadie} {et~al.}(2010){Abadie}, {Abbott}, {Abbott}, {Abernathy},
  {Accadia}, {Acernese}, {Adams}, {Adhikari}, {Ajith}, {Allen}, \&
  et~al.}]{LIGORates}
{Abadie}, J., {et~al.} 2010, CQGra, 27, 173001

\bibitem[{{Amati}(2006)}]{a06}
{Amati}, L. 2006, \mnras, 372, 233

\bibitem[{Amati {et~al.}(2013)Amati, Dichiara, Frontera, Guidorzi, Izzo, \&
  Valle}]{GCN15025}
Amati, L., {et~al.} 2013, GCN, 15025, 1

\bibitem[{Atwood {et~al.}(2009)Atwood, Abdo, Ackermann, Althouse, Anderson,
  Axelsson, Baldini, Ballet, Band, Barbiellini, Bartelt, Bastieri, Baughman,
  Bechtol, Bédérède, Bellardi, Bellazzini, Berenji, Bignami, Bisello,
  Bissaldi, Blandford, Bloom, Bogart, Bonamente, Bonnell, Borgland, Bouvier,
  Bregeon, Brez, Brigida, Bruel, Burnett, Busetto, Caliandro, Cameron, Caraveo,
  Carius, Carlson, Casandjian, Cavazzuti, Ceccanti, Cecchi, Charles, Chekhtman,
  Cheung, Chiang, Chipaux, Cillis, Ciprini, Claus, Cohen-Tanugi, Condamoor,
  Conrad, Corbet, Corucci, Costamante, Cutini, Davis, Decotigny, DeKlotz,
  Dermer, de~Angelis, Digel, do~Couto~e Silva, Drell, Dubois, Dumora, Edmonds,
  Fabiani, Farnier, Favuzzi, Flath, Fleury, Focke, Funk, Fusco, Gargano,
  Gasparrini, Gehrels, Gentit, Germani, Giebels, Giglietto, Giommi, Giordano,
  Glanzman, Godfrey, Grenier, Grondin, Grove, Guillemot, Guiriec, Haller,
  Harding, Hart, Hays, Healey, Hirayama, Hjalmarsdotter, Horn, Hughes,
  Jóhannesson, Johansson, Johnson, Johnson, Johnson, Johnson, Kamae, Katagiri,
  Kataoka, Kavelaars, Kawai, Kelly, Kerr, Klamra, Knödlseder, Kocian, Komin,
  Kuehn, Kuss, Landriu, Latronico, Lee, Lee, Lemoine-Goumard, Lionetto, Longo,
  Loparco, Lott, Lovellette, Lubrano, Madejski, Makeev, Marangelli, Massai,
  Mazziotta, McEnery, Menon, Meurer, Michelson, Minuti, Mirizzi, Mitthumsiri,
  Mizuno, Moiseev, Monte, Monzani, Moretti, Morselli, Moskalenko, Murgia,
  Nakamori, Nishino, Nolan, Norris, Nuss, Ohno, Ohsugi, Omodei, Orlando, Ormes,
  Paccagnella, Paneque, Panetta, Parent, Pearce, Pepe, Perazzo, Pesce-Rollins,
  Picozza, Pieri, Pinchera, Piron, Porter, Poupard, Rainò, Rando, Rapposelli,
  Razzano, Reimer, Reimer, Reposeur, Reyes, Ritz, Rochester, Rodriguez, Romani,
  Roth, Russell, Ryde, Sabatini, Sadrozinski, Sanchez, Sander, Sapozhnikov,
  Parkinson, Scargle, Schalk, Scolieri, Sgrò, Share, Shaw, Shimokawabe,
  Shrader, Sierpowska-Bartosik, Siskind, Smith, Smith, Spandre, Spinelli,
  Starck, Stephens, Strickman, Strong, Suson, Tajima, Takahashi, Takahashi,
  Tanaka, Tenze, Tether, Thayer, Thayer, Thompson, Tibaldo, Tibolla, Torres,
  Tosti, Tramacere, Turri, Usher, Vilchez, Vitale, Wang, Watters, Winer, Wood,
  Ylinen, \& Ziegler}]{LAT}
Atwood, W.~B., {et~al.} 2009, \apj, 697, 1071

\bibitem[{{Bloom} {et~al.}(2012){Bloom}, {Richards}, {Nugent}, {Quimby},
  {Kasliwal}, {Starr}, {Poznanski}, {Ofek}, {Cenko}, {Butler}, {Kulkarni},
  {Gal-Yam}, \& {Law}}]{BloomMachineLearning}
{Bloom}, J.~S., {et~al.} 2012, \pasp, 124, 1175

\bibitem[{{Brink} {et~al.}(2013){Brink}, {Richards}, {Poznanski}, {Bloom},
  {Rice}, {Negahban}, \& {Wainwright}}]{brp+12}
{Brink}, H., {et~al.} 2013, \mnras{} (in press)

\bibitem[{{Burrows} {et~al.}(2005){Burrows}, {Hill}, {Nousek}, {Kennea},
  {Wells}, {Osborne}, {Abbey}, {Beardmore}, {Mukerjee}, {Short}, {Chincarini},
  {Campana}, {Citterio}, {Moretti}, {Pagani}, {Tagliaferri}, {Giommi},
  {Capalbi}, {Tamburelli}, {Angelini}, {Cusumano}, {Br{\"a}uninger}, {Burkert},
  \& {Hartner}}]{bhn+05}
{Burrows}, D.~N., {et~al.} 2005, SSRv, 120, 165

\bibitem[{{Butler} {et~al.}(2007){Butler}, {Kocevski}, {Bloom}, \&
  {Curtis}}]{bkb+07}
{Butler}, N.~R., {et~al.} 2007, \apj, 671, 656

\bibitem[{{Campana} {et~al.}(2006){Campana}, {Mangano}, {Blustin}, {Brown},
  {Burrows}, {Chincarini}, {Cummings}, {Cusumano}, {Della Valle}, {Malesani},
  {M{\'e}sz{\'a}ros}, {Nousek}, {Page}, {Sakamoto}, {Waxman}, {Zhang}, {Dai},
  {Gehrels}, {Immler}, {Marshall}, {Mason}, {Moretti}, {O'Brien}, {Osborne},
  {Page}, {Romano}, {Roming}, {Tagliaferri}, {Cominsky}, {Giommi}, {Godet},
  {Kennea}, {Krimm}, {Angelini}, {Barthelmy}, {Boyd}, {Palmer}, {Wells}, \&
  {White}}]{cmb+06}
{Campana}, S., {et~al.} 2006, \nat, 442, 1008

\bibitem[{Cenko {et~al.}(2012)Cenko, Ofek, \& Nugent}]{GCN13489}
Cenko, S.~B., Ofek, E.~O., \& Nugent, P.~E. 2012, GCN, 13489, 1

\bibitem[{{Cenko} {et~al.}(2006){Cenko}, {Fox}, {Moon}, {Harrison}, {Kulkarni},
  {Henning}, {Guzman}, {Bonati}, {Smith}, {Thicksten}, {Doyle}, {Petrie},
  {Gal-Yam}, {Soderberg}, {Anagnostou}, \& {Laity}}]{P60Automation}
{Cenko}, S.~B., {et~al.} 2006, \pasp, 118, 1396

\bibitem[{Cheung {et~al.}(2013)Cheung, Vianello, Zhu, Racusin, Connaughton, \&
  Carpenter}]{GCN14971}
Cheung, T., {et~al.} 2013, GCN, 14971, 1

\bibitem[{Collazzi \& Connaughton(2013)}]{GCN14972}
Collazzi, A.~C., \& Connaughton, V. 2013, GCN, 14972, 1

\bibitem[{{Costa} {et~al.}(1997){Costa}, {Frontera}, {Heise}, {Feroci}, {in't
  Zand}, {Fiore}, {Cinti}, {Dal Fiume}, {Nicastro}, {Orlandini}, {Palazzi},
  {Rapisarda\#}, {Zavattini}, {Jager}, {Parmar}, {Owens}, {Molendi},
  {Cusumano}, {Maccarone}, {Giarrusso}, {Coletta}, {Antonelli}, {Giommi},
  {Muller}, {Piro}, \& {Butler}}]{GRBsHaveXrayAfterglows}
{Costa}, E., {et~al.} 1997, \nat, 387, 783

\bibitem[{{Dressler} {et~al.}(2011){Dressler}, {Bigelow}, {Hare}, {Sutin},
  {Thompson}, {Burley}, {Epps}, {Oemler}, {Bagish}, {Birk}, {Clardy},
  {Gunnels}, {Kelson}, {Shectman}, \& {Osip}}]{IMACS}
{Dressler}, A., {et~al.} 2011, \pasp, 123, 288

\bibitem[{{Evans} {et~al.}(2007){Evans}, {Beardmore}, {Page}, {Tyler},
  {Osborne}, {Goad}, {O'Brien}, {Vetere}, {Racusin}, {Morris}, {Burrows},
  {Capalbi}, {Perri}, {Gehrels}, \& {Romano}}]{ebp+07}
{Evans}, P.~A., {et~al.} 2007, \aap, 469, 379

\bibitem[{Fairhurst(2011)}]{FairhurstTriangulation}
Fairhurst, S. 2011, CQGra, 28, 105021

\bibitem[{{Frail} {et~al.}(1997){Frail}, {Kulkarni}, {Nicastro}, {Feroci}, \&
  {Taylor}}]{GRBsHaveRadioAfterglows}
{Frail}, D.~A., {et~al.} 1997, \nat, 389, 261

\bibitem[{{Gal-Yam} {et~al.}(2011){Gal-Yam}, {Kasliwal}, {Arcavi}, {Green},
  {Yaron}, {Ben-Ami}, {Xu}, {Sternberg}, {Quimby}, {Kulkarni}, {Ofek},
  {Walters}, {Nugent}, {Poznanski}, {Bloom}, {Cenko}, {Filippenko}, {Li},
  {Silverman}, {Walker}, {Sullivan}, {Maguire}, {Howell}, {Mazzali}, {Frail},
  {Bersier}, {James}, {Akerlof}, {Yuan}, {Law}, {Fox}, \&
  {Gehrels}}]{2011ApJ...736..159G}
{Gal-Yam}, A., {et~al.} 2011, \apj, 736, 159

\bibitem[{{Gehrels} {et~al.}(2004){Gehrels}, {Chincarini}, {Giommi}, {Mason},
  {Nousek}, {Wells}, {White}, {Barthelmy}, {Burrows}, {Cominsky}, {Hurley},
  {Marshall}, {M{\'e}sz{\'a}ros}, {Roming}, {Angelini}, {Barbier}, {Belloni},
  {Campana}, {Caraveo}, {Chester}, {Citterio}, {Cline}, {Cropper}, {Cummings},
  {Dean}, {Feigelson}, {Fenimore}, {Frail}, {Fruchter}, {Garmire}, {Gendreau},
  {Ghisellini}, {Greiner}, {Hill}, {Hunsberger}, {Krimm}, {Kulkarni}, {Kumar},
  {Lebrun}, {Lloyd-Ronning}, {Markwardt}, {Mattson}, {Mushotzky}, {Norris},
  {Osborne}, {Paczynski}, {Palmer}, {Park}, {Parsons}, {Paul}, {Rees},
  {Reynolds}, {Rhoads}, {Sasseen}, {Schaefer}, {Short}, {Smale}, {Smith},
  {Stella}, {Tagliaferri}, {Takahashi}, {Tashiro}, {Townsley}, {Tueller},
  {Turner}, {Vietri}, {Voges}, {Ward}, {Willingale}, {Zerbi}, \&
  {Zhang}}]{gcg+04}
{Gehrels}, N., {et~al.} 2004, \apj, 611, 1005

\bibitem[{{G{\'o}rski} {et~al.}(2005){G{\'o}rski}, {Hivon}, {Banday},
  {Wandelt}, {Hansen}, {Reinecke}, \& {Bartelmann}}]{HEALPix}
{G{\'o}rski}, K.~M., {et~al.} 2005, \apj, 622, 759

\bibitem[{Hurley {et~al.}(2013)Hurley, Goldsten, Connaughton, Briggs, Meegan,
  Pelassa, Golenetskii, Aptekar, Mazets, Pal'shin, Frederiks, Svinkin, Cline,
  von Kienlin, Zhang, Rau, Savchenko, Bozzo, \& Ferrigno}]{GCN14974}
Hurley, K., {et~al.} 2013, GCN, 14974, 1

\bibitem[{{Kann} {et~al.}(2010){Kann}, {Klose}, {Zhang}, {Malesani}, {Nakar},
  {Pozanenko}, {Wilson}, {Butler}, {Jakobsson}, {Schulze}, {Andreev},
  {Antonelli}, {Bikmaev}, {Biryukov}, {B{\"o}ttcher}, {Burenin}, {Castro
  Cer{\'o}n}, {Castro-Tirado}, {Chincarini}, {Cobb}, {Covino}, {D'Avanzo},
  {D'Elia}, {Della Valle}, {de Ugarte Postigo}, {Efimov}, {Ferrero}, {Fugazza},
  {Fynbo}, {G{\aa}lfalk}, {Grundahl}, {Gorosabel}, {Gupta}, {Guziy}, {Hafizov},
  {Hjorth}, {Holhjem}, {Ibrahimov}, {Im}, {Israel}, {Je{\'l}inek}, {Jensen},
  {Karimov}, {Khamitov}, {Kizilo{\v g}lu}, {Klunko}, {Kub{\'a}nek}, {Kutyrev},
  {Laursen}, {Levan}, {Mannucci}, {Martin}, {Mescheryakov}, {Mirabal},
  {Norris}, {Ovaldsen}, {Paraficz}, {Pavlenko}, {Piranomonte}, {Rossi},
  {Rumyantsev}, {Salinas}, {Sergeev}, {Sharapov}, {Sollerman}, {Stecklum},
  {Stella}, {Tagliaferri}, {Tanvir}, {Telting}, {Testa}, {Updike}, {Volnova},
  {Watson}, {Wiersema}, \& {Xu}}]{kkz+10}
{Kann}, D.~A., {et~al.} 2010, \apj, 720, 1513

\bibitem[{{Kelly} {et~al.}(2013){Kelly}, {Filippenko}, {Fox}, {Zheng}, \&
  {Clubb}}]{HostEnvironment}
{Kelly}, P.~L., {et~al.} 2013, \apjl, 775, L5

\bibitem[{{Kulkarni}(2012)}]{ZTF}
{Kulkarni}, S.~R. 2012, in Proceedings of the International Astronomical Union,
  Vol. 285, IAU Symposium, ed. E.~{Griffin}, R.~{Hanisch}, \& R.~{Seaman},
  55--61

\bibitem[{{Law} {et~al.}(2009){Law}, {Kulkarni}, {Dekany}, {Ofek}, {Quimby},
  {Nugent}, {Surace}, {Grillmair}, {Bloom}, {Kasliwal}, {Bildsten}, {Brown},
  {Cenko}, {Ciardi}, {Croner}, {Djorgovski}, {van Eyken}, {Filippenko}, {Fox},
  {Gal-Yam}, {Hale}, {Hamam}, {Helou}, {Henning}, {Howell}, {Jacobsen},
  {Laher}, {Mattingly}, {McKenna}, {Pickles}, {Poznanski}, {Rahmer}, {Rau},
  {Rosing}, {Shara}, {Smith}, {Starr}, {Sullivan}, {Velur}, {Walters}, \&
  {Zolkower}}]{PTF}
{Law}, N.~M., {et~al.} 2009, \pasp, 121, 1395

\bibitem[{{Li} \& {Paczy{\'n}ski}(1998)}]{kilonova}
{Li}, L.-X., \& {Paczy{\'n}ski}, B. 1998, \apjl, 507, L59

\bibitem[{{Meegan} {et~al.}(2009){Meegan}, {Lichti}, {Bhat}, {Bissaldi},
  {Briggs}, {Connaughton}, {Diehl}, {Fishman}, {Greiner}, {Hoover}, {van der
  Horst}, {von Kienlin}, {Kippen}, {Kouveliotou}, {McBreen}, {Paciesas},
  {Preece}, {Steinle}, {Wallace}, {Wilson}, \& {Wilson-Hodge}}]{FermiGBM}
{Meegan}, C., {et~al.} 2009, \apj, 702, 791

\bibitem[{{Meegan} {et~al.}(1992){Meegan}, {Fishman}, {Wilson}, {Horack},
  {Brock}, {Paciesas}, {Pendleton}, \& {Kouveliotou}}]{GRBsAreExtragalactic}
{Meegan}, C.~A., {et~al.} 1992, \nat, 355, 143

\bibitem[{{Mirabal} {et~al.}(2007){Mirabal}, {Halpern}, \&
  {O'Brien}}]{2007ApJ...661L.127M}
{Mirabal}, N., {Halpern}, J.~P., \& {O'Brien}, P.~T. 2007, \apjl, 661, L127

\bibitem[{Nissanke {et~al.}(2013)Nissanke, Kasliwal, \&
  Georgieva}]{NissankeKasliwalEMCounterparts}
Nissanke, S., Kasliwal, M., \& Georgieva, A. 2013, \apj, 767, 124

\bibitem[{{Nissanke} {et~al.}(2011){Nissanke}, {Sievers}, {Dalal}, \&
  {Holz}}]{NissankeLocalization}
{Nissanke}, S., {et~al.} 2011, \apj, 739, 99

\bibitem[{{Ofek} {et~al.}(2012){Ofek}, {Laher}, {Law}, {Surace}, {Levitan},
  {Sesar}, {Horesh}, {Poznanski}, {van Eyken}, {Kulkarni}, {Nugent},
  {Zolkower}, {Walters}, {Sullivan}, {Ag{\"u}eros}, {Bildsten}, {Bloom},
  {Cenko}, {Gal-Yam}, {Grillmair}, {Helou}, {Kasliwal}, \&
  {Quimby}}]{P48PhotometricCalibration}
{Ofek}, E.~O., {et~al.} 2012, \pasp, 124, 62

\bibitem[{{Perley} {et~al.}(2013){Perley}, {Cenko}, {Corsi}, {Tanvir}, {Levan},
  {Kann}, {Sonbas}, {Wiersema}, {Zheng}, {Zhao}, {Bai}, {Chang}, {Clubb},
  {Frail}, {Fruchter}, {G{\"o}{\u g}{\"u}{\c s}}, {Greiner}, {G{\"u}ver},
  {Horesh}, {Filippenko}, {Klose}, {Mao}, {Morgan}, {Schmidl}, {Stecklum},
  {Tanga}, {Wang}, \& {Xin}}]{pcc+13}
{Perley}, D.~A., {et~al.} 2013, \apj\ submitted (astro-ph/1307.4401)

\bibitem[{{Pian} {et~al.}(2000){Pian}, {Amati}, {Antonelli}, {Butler}, {Costa},
  {Cusumano}, {Danziger}, {Feroci}, {Fiore}, {Frontera}, {Giommi}, {Masetti},
  {Muller}, {Nicastro}, {Oosterbroek}, {Orlandini}, {Owens}, {Palazzi},
  {Parmar}, {Piro}, {in't Zand}, {Castro-Tirado}, {Coletta}, {Dal Fiume}, {Del
  Sordo}, {Heise}, {Soffitta}, \& {Torroni}}]{paa+00}
{Pian}, E., {et~al.} 2000, \apj, 536, 778

\bibitem[{{Rahmer} {et~al.}(2008){Rahmer}, {Smith}, {Velur}, {Hale}, {Law},
  {Bui}, {Petrie}, \& {Dekany}}]{rsv+08}
{Rahmer}, G., {et~al.} 2008, in \procspie, Vol. 7014, 70144Y--70144Y--12

\bibitem[{{Rhoads}(1997)}]{offaxis}
{Rhoads}, J.~E. 1997, \apjl, 487, L1

\bibitem[{{Robitaille} {et~al.}(2013){Robitaille}, {Tollerud}, {Greenfield},
  {Droettboom}, {Bray}, {Aldcroft}, {Davis}, {Ginsburg}, {Price-Whelan},
  {Kerzendorf}, {Conley}, {Crighton}, {Barbary}, {Muna}, {Ferguson},
  {Grollier}, {Parikh}, {Nair}, {Unther}, {Deil}, {Woillez}, {Conseil},
  {Kramer}, {Turner}, {Singer}, {Fox}, {Weaver}, {Zabalza}, {Edwards}, {Azalee
  Bostroem}, {Burke}, {Casey}, {Crawford}, {Dencheva}, {Ely}, {Jenness},
  {Labrie}, {Lian Lim}, {Pierfederici}, {Pontzen}, {Ptak}, {Refsdal},
  {Servillat}, \& {Streicher}}]{astropy}
{Robitaille}, T.~P., {et~al.} 2013, \aap, 558, A33

\bibitem[{{Sari} {et~al.}(1998){Sari}, {Piran}, \&
  {Narayan}}]{SariPiranNarayan98}
{Sari}, R., {Piran}, T., \& {Narayan}, R. 1998, \apjl, 497, L17

\bibitem[{{Savaglio} {et~al.}(2009){Savaglio}, {Glazebrook}, \& {Le
  Borgne}}]{sgl09}
{Savaglio}, S., {Glazebrook}, K., \& {Le Borgne}, D. 2009, \apj, 691, 182

\bibitem[{Schlafly \& Finkbeiner(2011)}]{SchlaflyExtinction}
Schlafly, E.~F., \& Finkbeiner, D.~P. 2011, \apj, 737, 103

\bibitem[{{Schulze} {et~al.}(2013){Schulze}, {Leloudas}, {Xu}, {Fynbo},
  {Geier}, \& {Jakobsson}}]{GCN14994}
{Schulze}, S., {et~al.} 2013, GCN, 14994, 1

\bibitem[{Singer {et~al.}(2013)Singer, Cenko, Kasliwal, Brown, Yaron, Bellm,
  Caudill, Tinyanont, Khatami, \& Weinstein}]{GCN14967}
Singer, L.~P., {et~al.} 2013, GCN, 14967, 1

\bibitem[{{Soderberg} {et~al.}(2004){Soderberg}, {Kulkarni}, {Berger}, {Fox},
  {Sako}, {Frail}, {Gal-Yam}, {Moon}, {Cenko}, {Yost}, {Phillips}, {Persson},
  {Freedman}, {Wyatt}, {Jayawardhana}, \& {Paulson}}]{2004Natur.430..648S}
{Soderberg}, A.~M., {et~al.} 2004, \nat, 430, 648

\bibitem[{{Soderberg} {et~al.}(2010){Soderberg}, {Chakraborti}, {Pignata},
  {Chevalier}, {Chandra}, {Ray}, {Wieringa}, {Copete}, {Chaplin},
  {Connaughton}, {Barthelmy}, {Bietenholz}, {Chugai}, {Stritzinger}, {Hamuy},
  {Fransson}, {Fox}, {Levesque}, {Grindlay}, {Challis}, {Foley}, {Kirshner},
  {Milne}, \& {Torres}}]{scp+10}
---. 2010, \nat, 463, 513

\bibitem[{{Tyson}(2002)}]{LSST}
{Tyson}, J.~A. 2002, in \procspie, ed. J.~A. {Tyson} \& S.~{Wolff}, Vol. 4836,
  10--20

\bibitem[{{van Paradijs} {et~al.}(1997){van Paradijs}, {Groot}, {Galama},
  {Kouveliotou}, {Strom}, {Telting}, {Rutten}, {Fishman}, {Meegan}, {Pettini},
  {Tanvir}, {Bloom}, {Pedersen}, {N{\o}rdgaard-Nielsen}, {Linden-V{\o}rnle},
  {Melnick}, {van der Steene}, {Bremer}, {Naber}, {Heise}, {in't Zand},
  {Costa}, {Feroci}, {Piro}, {Frontera}, {Zavattini}, {Nicastro}, {Palazzi},
  {Bennett}, {Hanlon}, \& {Parmar}}]{GRBsHaveOpticalAfterglows}
{van Paradijs}, J., {et~al.} 1997, \nat, 386, 686

\bibitem[{{Yaron} \& {Gal-Yam}(2012)}]{yg12}
{Yaron}, O., \& {Gal-Yam}, A. 2012, \pasp, 124, 668

\bibitem[{{Zhang} {et~al.}(2012){Zhang}, {Fan}, {Shen}, {Xu}, {Zhang}, {Wei},
  {Burrows}, {Zhang}, \& {Gehrels}}]{zfs+12}
{Zhang}, B.-B., {et~al.} 2012, \apj, 756, 190

\end{thebibliography}

\end{document}